\documentclass{bmcart}
\usepackage[utf8]{inputenc} 
\usepackage{graphicx}
\usepackage{amsfonts}
\usepackage{amsmath}
\usepackage{color}
\usepackage{algpseudocode}
\usepackage{algorithmicx}
\usepackage{algorithm}
\usepackage{verbatim}
\usepackage{nicefrac}

\startlocaldefs

\endlocaldefs

\begin{document}
\begin{frontmatter}

\begin{fmbox}
\dochead{Research}

\title{Uncovering nodes that spread information between communities in social networks}

\author[
   addressref={aff1},                   
   corref={aff1},                       
   noteref={n1},                        
   email={alexander.mantzaris@strath.ac.uk}   
]{\inits{AVM}\fnm{Alexander V.} \snm{Mantzaris}}
\address[id=aff1]{
  \orgname{Department of Mathematics and Statistics, University of Strathclyde}, 
  \street{26 Richmond Street},                     %
  \postcode{G1 1XH}                                
  \city{Glasgow},                              
  \cny{UK}                                    
}
\begin{artnotes}
\note[id=n1]{Equal contributor} 
\end{artnotes}

\end{fmbox}
\begin{abstractbox}

\begin{abstract} 

From many datasets gathered in online social networks, well defined community structures have been observed.
A large number of users participate in these networks and the size of the resulting graphs poses
computational challenges. 
There is a particular demand in identifying the nodes responsible for information flow between communities; for example, in temporal
Twitter networks edges between communities play a key role in propagating  spikes of activity
when the connectivity between communities is sparse and few edges exist 
between different clusters of nodes.
The new algorithm proposed here is aimed at revealing these key connections by measuring a node's vicinity to nodes of another community.
We look at the nodes which have edges in more than one community and the locality of nodes around them which influence the information
received and broadcasted to them.
The method relies on independent random walks of a chosen fixed number of steps, originating from nodes with edges in more than one community.
For the large networks that we have in mind, existing measures such as betweenness centrality are difficult to compute, even with recent methods that approximate the large number of 
operations required.
We therefore design an algorithm that scales up to the demand of current big data requirements and has the ability to harness parallel processing capabilities. 
The new algorithm is illustrated on synthetic data, where results can be judged carefully, and also on a real, large scale Twitter activity data, where new insights can be gained. 

\end{abstract}


\begin{keyword}
\kwd{Social Network Analysis}
\kwd{Community analysis}
\kwd{Twitter}
\kwd{Viral content}
\kwd{Community connectivity}
\kwd{Betweenness}
\kwd{Information diffusion}
\end{keyword}

\end{abstractbox}

\end{frontmatter}


\section*{Introduction}
\label{sec:intro}
Online social networks (OSNs) such as Facebook, LinkedIn and Twitter  have inspired a great amount of research.
Whether it is regarding their uses \cite{skeels2009social} in different aspects of our daily lives
or on how a  important scientific breakthrough 
can  spread around the world \cite{de2013anatomy}.
These networks can be very large, for example Facebook currently  holds around 1 billion user accounts.
Despite the obvious computational challenges, 
analysis of these large datasets provides the opportunity to test hypothesis
about human  social behavior on an unprecedented scale, and hence  
to reveal deeper understandings of human  social behavior \cite{mcalexander2002building}.
Furthermore,  commercial, government and charitable  enterprises 
can 
utilize the networks to inform campaigning, advertising and promotion. 
Hence,  there is great potential impact for improvements in the analytical tools 
designed for analysing social networks.

Within the OSNs generated by  users, community structures form naturally, and research into their 
detection is very active \cite{leskovec2009community,fortunato2010community}. These developments in community detection have produced a diverse set of methods which are at our disposal. 
Run times of the algorithms are a major concern, 
and current datasets
can be too large for many of the algorithms available. 
One approach  to deal with the size is by using network samples; for example, 
\cite{ferrara2012large} analyzes community structure
in a subset of millions of nodes taken from Facebook. 
However, for the types of  effects that span over the entirety of the networks, we wish to avoid sampling and deal with complete networks.

Communities in OSNs can emerge for many reasons. A key driver can be homophily \cite{mcpherson2001birds}, where some underlying similarity between users in a community leads to a higher number of edges between these users than with users in a different community.
\cite{matsuo2009community} investigates homophily formation and evolution in a online social buyers setting. 
Here, 
a community builds trust and supports the activity of online purchases, which is the motivation for more in depth research
into the nature of the inter-community connections. Companies have an interest
in their brand identity within OSN communities, as users now have the ability to
broadcast brand information to many other users within their social reach. 
Although not based on data 
from OSNs, \cite{mcalexander2002building} discusses the
attributes that users utilize their associations, and how companies should work to cultivate
their brand presence with customers. The authors also raise many interesting questions concerning
the dynamic elements of brand presence which are relevant to our work.

It is also interesting to elaborate on how distinct communities are 
brought together to create large connected graphs. Without the connectivity between dense communities,
isolated components would not support many of the fascinating phenomena 
that have been observed, notably the hugely influential small world effect \cite{travers1969experimental},
where there exist surprisingly short paths between members of the network located in different communities.
By definition, 
the density of those edges connecting communities is less than the density of edges within communities. The sparsity of the 
inter-community connectivity is the basis for community separation quality measures such as the modularity index, \cite{newman2004finding}.
The relatively low number of these edges connecting communities together gives them special importance as they are critical for the 
graph's connectivity. A recent study explores this network feature using examples from brain connectivity, \cite{clune2013evolutionary}, 
concluding that connection costs can explain these modular networks. 
For the applications in OSNs, where companies seek 
to harness the power of internet advertising, nodes which offer community traversal connections are critical targets
\cite{subramani2003knowledge}. 
The aim of our work is therefore to give a simple and scalable methodology for defining and 
discovering this type of key structure. 
To be concrete,  
for the remainder of this paper an edge connecting two different communities will be referred
to as a \emph{boundary edge} and the nodes on either side of these edges as \emph{boundary nodes}.

It is important to  have in mind that the edges are created
by means of an information exchange where nodes that receive  content independently decide on whether to 
repeat this received information to its follower node set, and in a future time step that can include nodes that were 
previously not included in the sharing of this content for whatever reasons or constraints might exist.
For content to spread throughout the network this decision to repeat the content must be consistently 
agreed on independently. The number of times this must occur is  increased when there is a large number of communities and only a few boundary nodes acting as regulators for the content to become \emph{viral}. The term viral usually assumes
that a large portion of the nodes in network are aware of a piece of information or content regardless of the specific niche community they may belong to.
Viral activity can be identified through conversation spikes, or cascades, as 
users share a common piece of content in a short amount of time. 

In general the content users would classify as \emph{news} has many examples of viral spreading of content.
Twitter is sometimes considered to be a news source, with \cite{kwak2010twitter} counting
at least 85\% of tweets being related to headline news.
Much of which  includes news   of commercial interest and opens possibilities for real time engagement.
Real time monitoring of these events being discussed is therefore essential for automated 
engagement. With spikes in topics lasting in the magnitude of minutes,
the run time of an algorithm should be reduced as much as possible and the ability 
of the algorithm to utilize 
the hardware of multiple processors is highly desirable . 
The work of \cite{weng2011event,nichols2012summarizing} discusses this real time monitoring of events and gives a number of case studies, comparing techniques for spike detection.
Our work has a slightly different emphasis, since we aim to detect nodes and edges that 
facilitate propagation of information, and hence would be natural candidates for monitoring
and intervention.

To introduce notation and background, we consider a 
 a graph $G=(V,E)$, with $N=|V|$ number of nodes and $M=|E|$ as the number of edges.
The standard 
centrality measure most relevant to our work is 
betweenness (shortest path betweenness), \cite{freeman1977set}.
For a node $v$, this measure is defined as 
\begin{equation}
b_v = \sum_{i\neq j\neq v}\frac{\sigma_{i,j}\left( v \right)}{\sigma_{i,j}}
\end{equation}
where, $\sigma_{i,j}$ counts the total number of shortest paths between $i$ and $j$, and $\sigma_{i,j}\left( v \right)$ counts how many of these 
pass through node $v$. 
Hence, $b_v$ gives an indication for the amount of potential control or influence node $v$ has on the information flow between all other nodes in the network.
Computing this measure straightforwardly for each node requires a large number of operations, $\Theta\left( N^3 \right)$, leading to a run time that 
is impractical for large networks. Using Brande's algorithm \cite{brandes2001faster} a 
complexity of $\Omega\left(M\times N\right)$ is possible, which is still 
time consuming for the networks with millions of nodes. 

The strict assumption that information flows along shortest paths (geodesics) is not 
always appropriate, as discussed, for example,  by Newman \cite{newman2005measure}, who  proposes a \emph{random walk betweenness} measure computed using matrix methods.
An important criticism of the geodesic viewpoint, which also motivates the random walk alternative, is that when passing messages to target nodes, typical 
users do not have  the global network information and  hence may not be aware of
the shortest paths between pairs of nodes to be able to place them along the correct route. 
The runtime for this \emph{random walk betweenness} measure is $\Omega \left(\left(M+N\right)N^2 \right)$ and the algorithm requires matrix inversions.
We also note that these two betweenness measures  above are designed for static 
networks, and changes in the size of communities over time
can affect the distribution of the betweenness values amongst the nodes.

\section*{Methodology}
\label{sec:methodology}
Given networks arising from online social media, there are many cases where rich community structure is observed.
The edges connecting these separately clustered groups of nodes are referred to as boundary nodes here, and those edges connecting them
as boundary edges. 
In this section our algorithm for measuring the boundary node proximity is described. 
The goal is to be able to rank nodes in a network according to their ability to influence nodes across different communities by the 
information (content) they exchange. This will reveal the boundary nodes, which play a key role in exchanging information between different communities, and those nodes surrounding them in their local vicinity.
The algorithm is based on the premise that 
information travels via a random walk rather than through a shortest path route.

An adjacency matrix $A$ of dimension $N$ will be used to represent the original network, where $A_{i,j}=1$ when there is an edge between nodes $i$ and $j$.
Once the network has  been decomposed into its connected subcomponents and the community labelling has been assigned, 
the set of boundary edges, connecting two nodes $\left(i,j\right)$ belonging to different communities, 
can then be defined as:
\begin{equation}
\label{eq:boundaryEdgeDef}
\mathbf{W}_{i,j} = \{\left(i,j\right):i\in C_1,j \not\in C_1,i \not\in C_2,j\in C_2\}.
\end{equation}
Here $\left(C_1,C_2\right)$ are two communities belonging to the 
list of community labels, $\mathbf{C}$, in the graph. 
We assume that the number of community labels will be much less than the number
of nodes, $|\mathbf{C}| \ll N$. From the boundary edge set $\mathbf{W}$, the boundary nodes  $\mathbf{B}$, can be found.
Due to the typical sparsity of the community connectivity, 
the number of boundary nodes will be much less than the total number of nodes,
$|\mathbf{B}| \ll N$.

The algorithm proposed here iterates through the boundary node set and performs a set of independent truncated random walkers originating at each boundary node, until convergence is reached in the distribution of visits to the nodes in the vicinity of each boundary node. It is the counts of the visits from the random walkers to the boundary nodes and nodes in their vicinity which allows a ranking in terms of being able to influence another community by spread of content.
The description of the algorithm is summarized in steps 1 to 4 of the outline given in  
Table~\ref{alg:outline}.
In the first step of the algorithm
the set of connected graphs is extracted from the original graph of the network, $\textbf{G}$,
using breadth first search (BFS). There exist many efficient methods for performing BFS and in
\cite{reingold2008undirected} it is shown that the step of acquiring the set of connected components 
can be performed in $\Omega \left( \text{log}N\right)$. The second step 
is to label each node according to the community it belongs to.
There is a wide selection of algorithms for obtaining the community structure, \cite{fortunato2010community,lancichinetti2009community}. In our work, we use the
Louvain method of \cite{blondel2008fast}. The run time of the Louvain algorithm is 
$\Omega\left( N\text{log}N \right)$, and there is an efficient implementation available. 
Tests run with this method report working with millions of nodes under 2 minutes on a standard PC. 
It is a greedy algorithm using the modularity index, \cite{clauset2004finding}, as an optimisation criteria, 
which has the benefit that the number of iterations taken by the algorithm can be controlled to some extent by examining
the value of the change in the index per iteration.
The third step extracts the set of boundary nodes which control the information exchange between different communities as they are the only nodes that connect directly to nodes in a different community.
 
The final step in Table~\ref{alg:outline} runs a number of i.i.d.\ random walkers from each boundary node until a convergence criterion is satisfied based on the number of visits to the nodes in the network. The number of visits
to each node is counted and is a measure of ability to disseminate information across boundary edges and influence
different communities.
Steps 3 and 4 are described in more detail in the next subsection. 
This algorithm can be referred to as the boundary vicinity algorithm (BVA).

\subsection*{Boundary node analysis}
\label{subsec:Bnodeanalysis}

To obtain the boundary nodes/edges
of a connected graph as defined in (\ref{eq:boundaryEdgeDef}), we use 
the vector of community labellings for the set of nodes in the network $\mathbf{C}$ and look for adjacent nodes with different
community labellings. 
With an edge list of the adjacency matrix of the graph, $L$, each edge is represented as a row number and a column number in this two column matrix. 
Where the two labels differ on a row in this edge list, a boundary edge has been detected;
\begin{equation}
\label{eq:bridgeEdge}
\mathbf{W} = \{\mathbf{C}\left(L\left(s,1\right)\right) \neq \mathbf{C}\left(L\left(s,2\right)\right)\}.
\end{equation}

 The adjacency matrix of the community specific graph is the
 matrix $A_{C}(i,j)$, where 
\begin{equation}
\label{eq:comAdj}
A_{C_l :i,j} = \left\{ 
  \begin{array}{l l}
    1 & \quad \text{$A_{i,j}\times \delta \left(C_{l}(i), C_{l}(j)\right)= 1$}\\
    0 & \quad \text{otherwise}.
  \end{array} \right.
\end{equation}
Here the $\delta(\cdot,\cdot)$ is the kronecker delta where the value of one is given when both inputs are equal. 
To obtain these matrices it is not a requirement to iterate through each element. This will be clarified below.
With the community adjacency matrices for each community label $A_{C}$ and the boundary nodes belonging to the network $\mathbf{B}$
we can iterate through the nodes  of $\mathbf{B}$ and run the series of random walkers localised on each boundary node
and confined to each $A_{C}$.

The random walks used to measure the ability for nodes to influence and affect the boundary nodes have a fixed number of steps.
For the walks to represent the localized region of these nodes, $\mathbf{B}$, the walks cannot be given an excessively large
length as this would dilute the importance of nodes closer to the boundary nodes.
The Barab\'{a}si–Albert model \cite{Barabasi99,albert2002statistical} 
uses the mechanism of preferential attachment to 
reproduce the growth characteristics of many networks. The average path length for these networks is $\nicefrac{\text{log}(N)}{\text{log}(\text{log}(N))}$, where we assume that 
the ceiling of the value is taken.
We use this value as a baseline is deciding 
the number of random walk steps that must be taken before a piece of information loses the consistency and relevance of the original content.

Based on this idea, 
Figure\ref{fig:bnvALG} displays the pseudocode of the algorithm for measuring the boundary node proximities.
The first input is the data structure for the connectivity of the nodes in the network. The second
input is $walknum$, which is the number of random walkers
that are started from each boundary node before convergence is tested. The third input, $stepnum$, is the number of steps/node traversals taken by each random walker.
The vector $visitCounts$ holds the number of visits by random walkers which are made to each node in the network
throughout the algorithm. As a variation this vector could be made into a sparse matrix where each row is the 
contribution of an i.i.d.\ walker, so that more information from the walks can be found.
The function call here, $\text{connected\_components}(G)$, is to the breadth-first-search algorithm which
produces a list of the elements in $G$ which are connected to each other.
Using the identifiers of connected graph membership, a list of the connected
components is produced, $\mathbf{G}$. In the loop of the elements of $\mathbf{G}$, $g$ is a connected network.
Here the function call $\text{community}(g)$, is to the community detection algorithm of choice (in this work the Louvain algorithm is used). This
returns the community membership labels $\mathbf{C}$ for the networks and modularity index $Q$. If the modularity index is not larger than
a certain threshold of choice, then it is considered that $g$ has no evident community structure and therefore no boundary nodes and the loop continues to the next  connected graph component $g$.
The boundary edges, $\mathbf{W}$, can be extracted by including only edges which connect different community labels.
We obtain the adjacency matrices for each community label by including only the nodes in each community label in a separate adjacency matrix.
For each unique $node$ in the list of $\mathbf{W}$ 
the algorithm then proceeds through the standard method of a set of random walks on the adjacency matrix. After
the completion of $walknum$ number of walks, the trajectories are tested for convergence using the potential scale reduction factor (PSRF) of
\cite{gelman1992inference}. If the set of trajectories have not converged, more walkers are computed until the required amount of 
convergence is achieved. Upon convergence, the values counts of visits to each node by the walkers is normalized to remove the 
effects of more walks due to lack of convergence. 
The results are scaled according to the 
relative size of the community in relation to the whole graph because of the impact it may have on the large scale of the information flow, so that the nodes in larger communities deliver a larger impact than small ones.

The run time of this algorithm is dominated by the community detection phase.
Due to  the boundary node set being much smaller in size than  the number of nodes, the loops required
to iterate through them and perform the random walks will typically cost less than $N$.

The last steps of the  algorithm can  naturally be parallelized by running the i.i.d.\ random walkers on separate
processors at the same time. After they have completed their walks the trajectories can then be monitored for convergence. 
The community detection component can also be parallelized by using the method of \cite{martelot2013fast} resulting in a completely parallelizable algorithm.

\section*{Results}
\label{sec:Results}
Here the results of using the algorithm on  synthetic datasets and a real dataset are shown.
One synthetic network used for testing is a set of random Erd{\"o}s-R{\'e}nyi (ER) graphs produced and  connected together to from a connected network by choosing randomly members to act as boundary nodes.
The other synthetic network is produced from connecting independent communities graphs produced using
 preferential attachment.
The well known Zachary Karate club dataset \cite{zachary1977information} is analyzed and presented. The Enron email dataset \cite{chapanond2005graph} is 
also analysed utilizing the valuable semantic data associated with the nodes to show the qualitative validity of the algorithm.
Lastly a new dataset collected from monitoring a Twitter hashtag is presented where the volume of 
Tweets and the volume of  boundary nodes are presented against a random set to show the importance that these nodes
have in spreading information throughout the network.

Figure\ref{fig:ER1} shows the results of using the boundary vicinity algorithm (BVA) and calculating betweenness on a synthetically produced 
network. Three communities were generated independently with the ER model and then a set of random
nodes (26 here) were selected from these communities to be connected to a different community. These 
selected nodes become the \emph{boundary nodes} in the network. 
There are six subfigures labelled a)-f), where
a) and b) show the normalized values from the algorithms (yaxis)
given to each node in the network (xaxis). Subfigure a) for the boundary vicinity algorithm has a more evenly spread
distribution across the nodes than what betweenness produces in subfigure b). We can see that betweenness gives almost
absolute importance to the nodes on the boundary with little emphasis for the nodes in the vicinity of those boundary nodes. 
Subfigures c) and d) display the networks with the vertices scaled according to the boundary vicinity measure and 
betweenness respectively. In c) we can see the neighbouring nodes of the boundary scaled as well.
Subfigure e) counts the proportion of overlap in the ranking between BVA and betweenness for an increasing number of 
nodes. We can see that both algorithms have almost complete overlap in choosing the top 26 nodes but differ in the 
order for the subsequent nodes. Subfigure f) shows a scatter plot of the values for all the
nodes with both algorithms. We can see how the top ranking nodes are clearly distinct from the bulk of the network and how BVA produces
a greater variance for nodes not in the boundary set.

In figure\ref{fig:PA1} 3 communities are produced using 
the  Barab\'{a}si–Albert model \cite{Barabasi99,albert2002statistical} algorithm of preferential attachment and then these communities
are   connected by choosing nodes uniformly from each group.
The same format as with the previous figure is used. In the first row of subplots, a) and b), we can see again that there is a wider
distribution in the scores for the nodes with the BVA algorithm on non-boundary nodes. 
In subfigures c) and d) we visualise the networks with the nodes scaled according to the BVA and betweenness respectively.
We can see that the highest degree nodes
which are central to the community they belong to are scaled and highlighted in both cases.
 A critical difference
is that the boundary nodes at the top which receive a large score with BVA but are given minimal importance 
with betweenness. With betweenness the role of these nodes is redundant given alternative routes through nodes with higher 
degree and direct connections to many nodes in the community. 
In the effort to inspire cross pollination of communities with promoted content, the ability to saturate a user
with fewer connections may be advantageous, and worth considering because they may be influenced more easily. 
In e) we look at the overlap proportion of the ranking between nodes for a number of nodes in both algorithms. 
We can see the local peak of the number of
overlaps for more nodes than the number of boundary nodes. This is because the structure of the network includes 
nodes in the vicinity of the boundary which lay on the shortest paths to other nodes in the community.
In the last subfigure we can see the scatter plot of the BVA values and betweenness. The ranking of the algorithms may 
be more similar to each other than with the ER communities connected but the distribution is much more narrow for betweenness
in this case, highlighting the few boundary nodes that are also core to the communities.

Figure~\ref{fig:karate} shows the results of using BVA on the Zachary karate club dataset. In subplot a) the network
is visualised and the
vertices are scaled according to normalised scores given by the BVA algorithm. 
The central members of the communities are given large
values as are the boundary nodes since they are within the vicinity of the boundary. 
In subfigure b) the overlap of the rankings with BVA and betweenness is shown for the number of nodes included,
and as with the previous 
two figures the overlap for both methods peaks when including the top number of nodes which corresponds to the
number of boundary nodes.

 When analyzing the Enron email dataset, a subset of the nodes are included where the
position in the company is known. BVA and betweenness scores are calculated for 
each of the nodes in the network and the top ten nodes for which their roles are known are
compared. BVA selects 3 vice presidents, 1 CEO, 2 managers, 2 traders, and 2 employees to be in the top ten.
Betweenness selects 1 vice president, 1 managing director, 2 managers, 1 director of trading, 
2 traders, 1 secretary and 2 employees. The list provided by BVA contains more company members with higher positions
than by betweenness. This may not be always the case, but it does show that the features of the network extracted
by BVA captures importance in the node placements.

Figure\ref{fig:temporal1} shows the Twitter activity of a TV show \emph{FearneHolly} which is monitored in real time using
the paid Twitter api service that does not deliver such a limited subset of the Tweets being sent regarding the hashtag as does the free service. 
We look at the Tweet volumes over time for this topic and plot them in the bottom subplot of the figure. We can see
a single dominating conversation intensity spike. 
We wish to test the necessity and importance  of the boundary node set, $\mathbf{W}$,  in facilitating the productions of these spikes of conversation activity in Twitter.
Since these dominant spikes observed in Twitter stand out so much from smaller oscillations it is assumed here that
it is due to the conversation taking place across the entire network and not confined to the locality of clustered nodes in a
particular community. 
The  boundary nodes of the network are found and over time their number counted at each time point is shown in the blue line in the first subplot. 
We see a single dominant spike for the number of boundary nodes  in the same region as that for the total conversation intensity in the bottom subplot. There is a need to test whether the boundary node increase at the same time as the total volume indicates that they provided vital routes for content to spread or is their number only a consequence of the overall activity of the nodes uniformly over the network and not in any way dependent on the presence of the boundary nodes. 
To investigate this, we select a random set of nodes in the network of equal size to that of the 
boundary node set and look for spikes in the volume of communication in this random set.
The random set of nodes we count at each time point produces an extra spike which is a false positive because it is not detectable in the bottom subplot
of the total Tweet activity. When the boundary nodes are active the random node selection and the total volume 
produce a spike reinforcing the idea that the nodes connecting communities are essential as a prerequisite for the 
total number of nodes to spike in conversation activity. The boundary node vicinity algorithm using this information
 can provide a evidence for spike detection and excitation (the dataset can be provided by contacting the author).


\section*{Discussion}
\label{sec:disc}
The work presented here gives an efficient algorithm for ranking the ability of nodes in a network, with community structure, to spread information
between clusters. Previously proposed methods impose large computational difficulties or are not based on principles which 
realistically model how information across the communities can spread. Focusing attention on these
boundary nodes in a network can be critical
for monitoring whether content may reach the point of becoming \emph{viral}. 
In practice nodes not all nodes may be directly influenceable, but may be indirectly by the nodes in their local 
vicinity. The boundary vicinity algorithm acknowledges nodes that may be placed in such a position to
have more or less influence on content leaving or entering a community of nodes in network.

A strength of this boundary vicinity algorithm is that it combines the power of community detection algorithms with the use of random walkers to assist in the process
of investigating the range of influence of the boundary nodes. The results show that 
this algorithm is comparable with 
 betweenness centrality without the requirement for full the maturity of a network to be visible.
In situations where the observed connectivity is changing, analysing the network in sections based on a community structure is an approach to provide more consistent results over time. 
The algorithm has a single tuning parameter which determines the number of steps a random walker takes from the boundary nodes.
Using a fraction of the average path length for networks constructed with the Barab\'{a}si–Albert model has given stable results in our experiments. 

Measures such as betweenness can provide a set of optimal targets for spreading  
content along shortest path routes throughout the complete connected network. This task ignores the challenges
that might be faced which attempting to promote activity in the critical set of nodes which lay on the 
boundary of the communities making up the complete network. A list of the nodes which 
are best positioned to quickly spread a piece of content does not address many of the practical 
challenges in inspiring activity as a non-invasive influencer.
Assessing the vicinity of the influencers for the boundary nodes gives reasonable subset for which 
attention must be given to ensure that cross pollination between clustered sets of nodes can occur.

Overall, the proposed  algorithm has the potential to quickly
handle the task of analysis with an online stream of large datasets. In particular
 real time event monitoring in environments such as Twitter 
where topic discussions can grow and decay rapidly, this is especially important.
With the goal of spreading the content as far as possible the boundary nodes, and those nodes
in its close vicinity in a community must 
be targeted, which is at the core of this method proposed here.

\begin{backmatter}
\section*{Acknowledgements}
Thanks is given to Peter Laflin for providing feedback over the course of this work, and to Bloom Agency, Leeds, for supplying anonymised Twitter data.

This work was performed as part of the
Mathematics of Large Technological Evolving Networks
(MOLTEN) project, which is supported by the Engineering and Physical Sciences Research Council and the Research
Councils UK Digital Economy programme, with grant ref. EP/I016058/1, and the
support of the Univeristy of Strathclyde with Bloom Agency for 
the  follow-on support from the Impact Acceleration Account.

\bibliographystyle{bmc-mathphys} 
\bibliography{mrefs-v7}

\section*{Figures}

\begin{figure}[h!]
\begin{algorithmic}
	\Procedure{BNV}{$G$,walknum,stepnum}
	\State visitCounts = $\leftarrow$ zeros(1,$N$) 
    	\State $ids$ $\leftarrow$ connected\_components($G$) 
	\ForAll{ id $\in ids$}
		\State $\mathbf{G}_i \leftarrow G(id,id)$ \Comment{$G_i$ is a connected graph}
	\EndFor		
	\ForAll{ $g \in \mathbf{G}$}
		\State $(\mathbf{C},Q) \leftarrow$ community($g$) \Comment{infer community structure}
		\If{$Q<$ threshold}
			\State continue
		\EndIf		
		\State $\mathbf{W}$  =  $g(i,j)\times(1 - \delta(C_{i} \neq C_{j}))$\Comment{boundary edge list}
		\State $\mathbf{A}_C = $community\_mask$(g,C)$
		\ForAll{$node \in \mathbf{W}$}						
			\State $A$ = $\mathbf{A}_{node \in C}$		
			\State present\_node = $node$
			\State $w = 0$
			\State walks = null
			\While{$w < $ walknum}\Comment{i.i.d walkers}				        		 	
        			\State pWeightsVec($node$) = 1
				\State $s = 0$
				\While{$s < $ stepnum}\Comment{random walker}					
        				\State $node$ = randomStateTransition($A$,$node$)
					\State pWeightsVec($node$) = 1 + pWeightsVec($node$)
					\State $s = s+1$
				\EndWhile				
				walks = append(walks, pWeightsVec];
				\State $w = w+1$
				\If{$w = $walknum} 
					\State psrf = GelmanRubinDiagnostic(walks);
					\If{psrf $< 0.95$ | psrf$ > 1.05$}
						\State $w = 0$ \Comment{more walks for convergence}
					\Else
						\State pWeights = scaleCommunityWeights(pWeightsVec)
						\State visitCounts = append(visitCounts,pWeightsVec) 
					\EndIf
				\EndIf
      			\EndWhile						 
		\EndFor
	\EndFor           
      \State \textbf{return} visitCounts
    \EndProcedure
  \end{algorithmic}
\caption{\label{fig:bnvALG} Boundary vicinity algorithm (BVA).  }
\end{figure}

\begin{figure}
\includegraphics[scale=0.35]{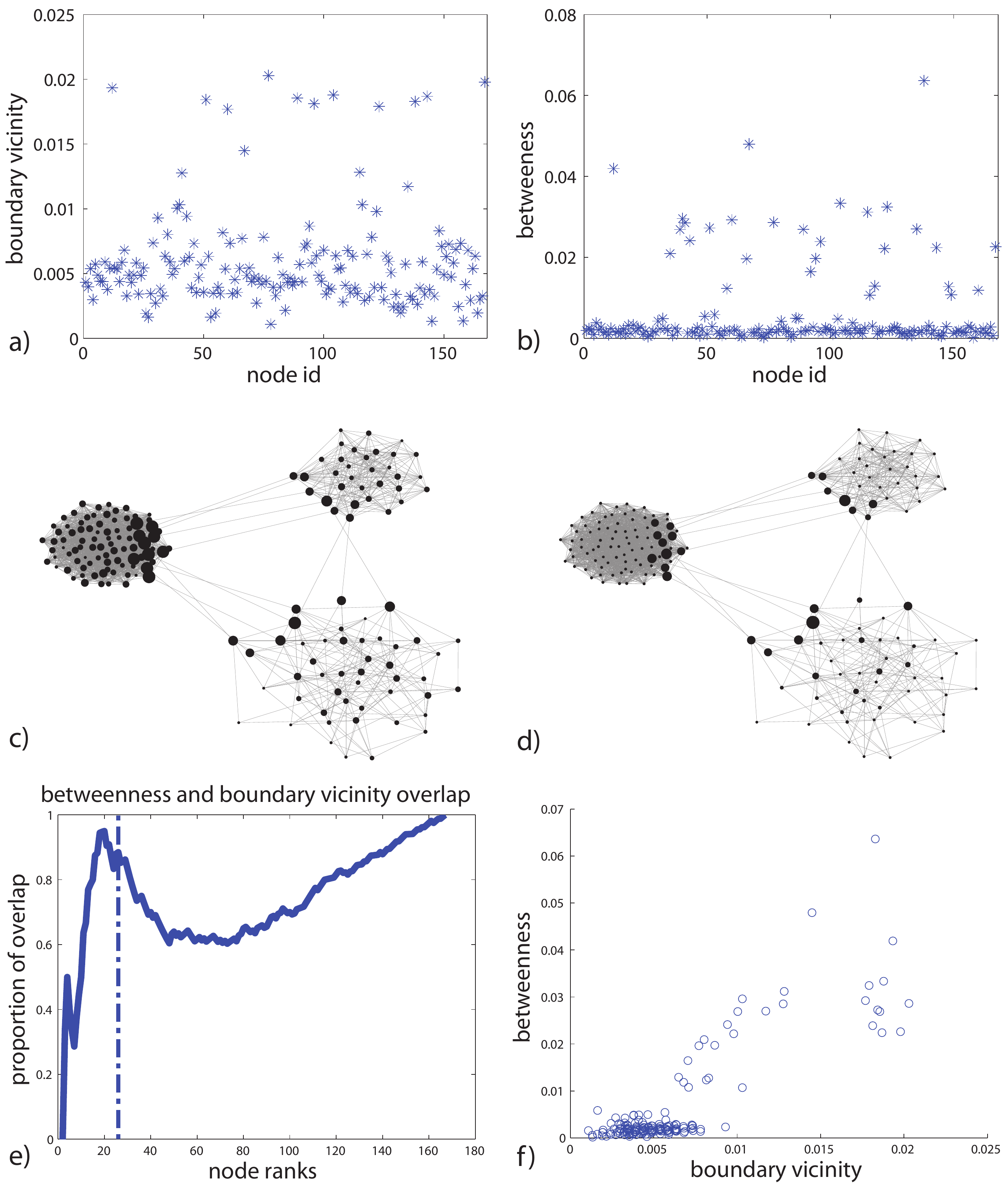} 
\caption{\label{fig:ER1}\csentence{3 ER communities connected and analysed with BVA and betweenness} 
The first row of plots show the scores given to the node IDs with the boundary vicinity algorithm and betweenness respectively. The second row of plots is the network visualizations with the nodes scaled in size according the normalized 
scores from the boundary vicinity algorithm and betweenness respectively. Subplot e) shows the proportion of overlap 
between the top ranking nodes from BVA and betweenness for different sizes of the ranking size. Subplot f) is a scatter
plot of the BVA and betweennes values for each node.}

\end{figure}

\begin{figure}
\begin{center}
\includegraphics[scale=0.35]{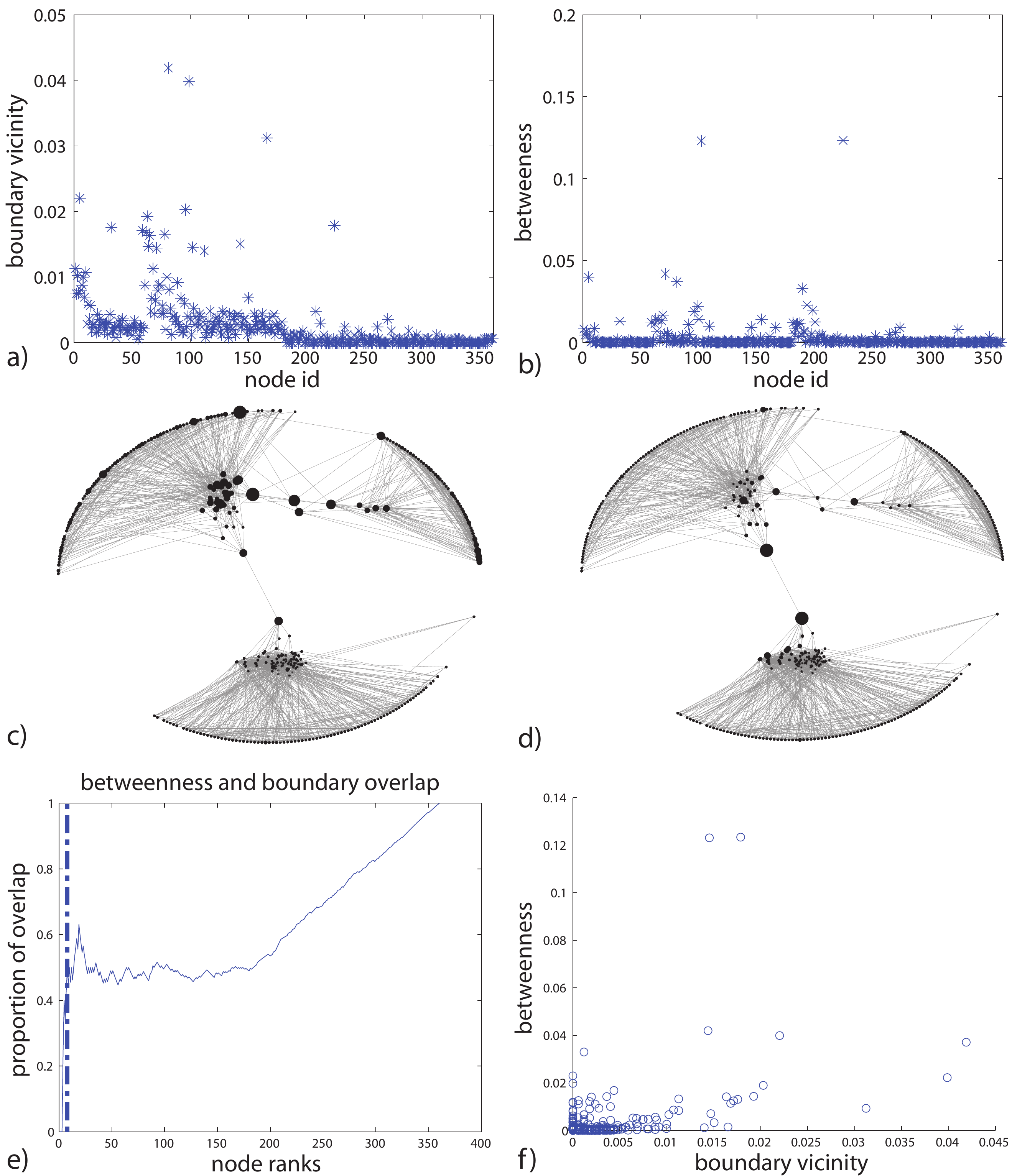} 
\caption{\label{fig:PA1}\csentence{3 communities produced with preferential attachment connected and analysed with BVA and betweenness} 
The first row of plots show the scores given to the node IDs with the boundary vicinity algorithm and betweenness respectively. The second row of plots is the network visualizations with the nodes scaled in size according the normalized 
scores from the boundary vicinity algorithm and betweenness respectively. Subplot e) shows the proportion of overlap 
between the top ranking nodes from BVA and betweenness for different sizes of the ranking size. Subplot f) is a scatter
plot of the BVA and betweennes values for each node.}
\end{center}
\end{figure}

\begin{figure}
\begin{center}
\includegraphics[scale=0.5]{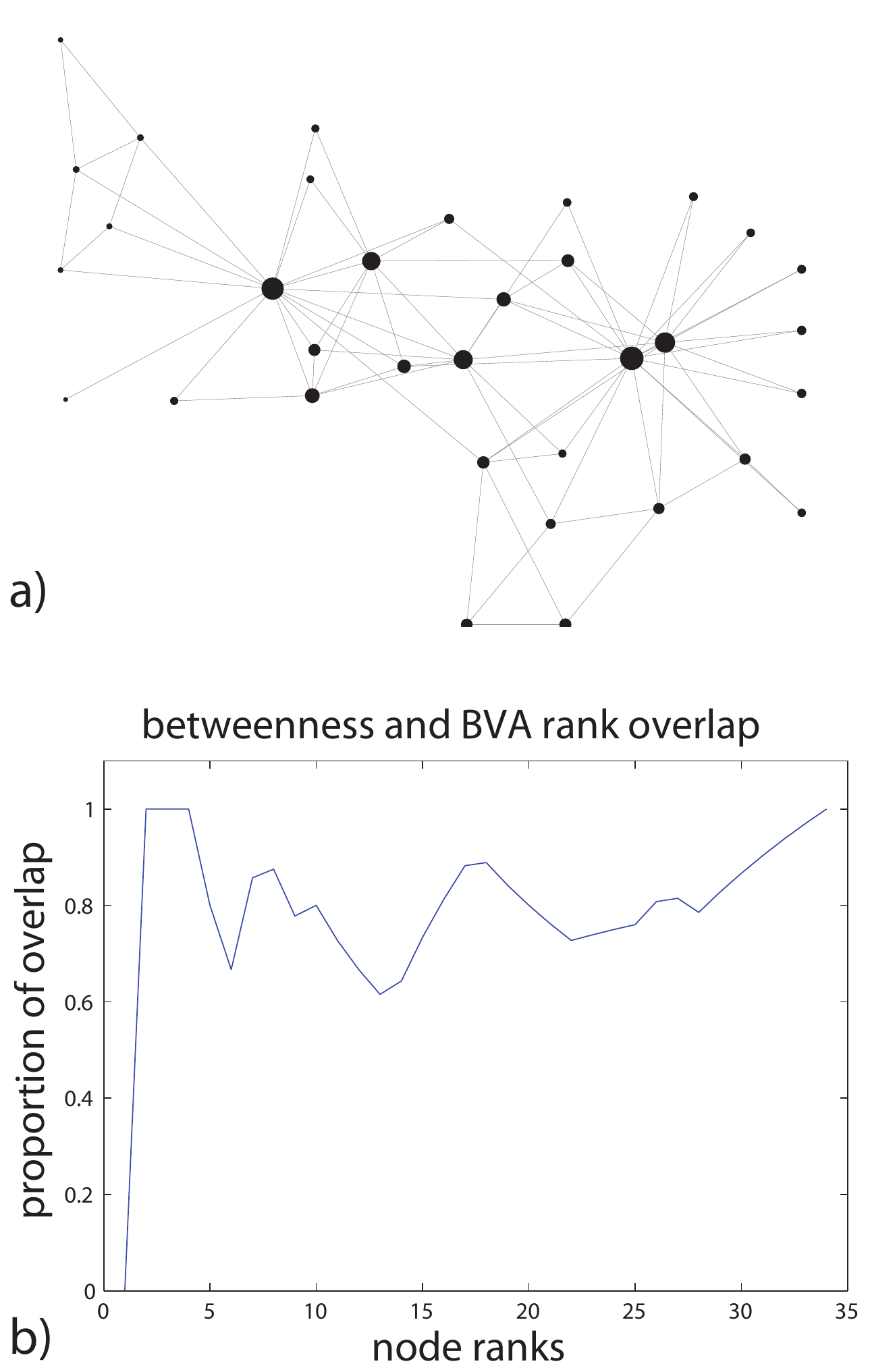}
\caption{\label{fig:karate}\csentence{Using the boundary vicinity algorithm on the Zachary Karate club dataset} 
Subfigure a) shows the network vertices scaled according to the normalised values given by BVA.
Subfigure b) shows the rank proportion overlap of BVA and betweenness for a number of different rank sizes.}
\end{center}
\end{figure}

\begin{figure}
\begin{center}
\includegraphics[scale=0.35]{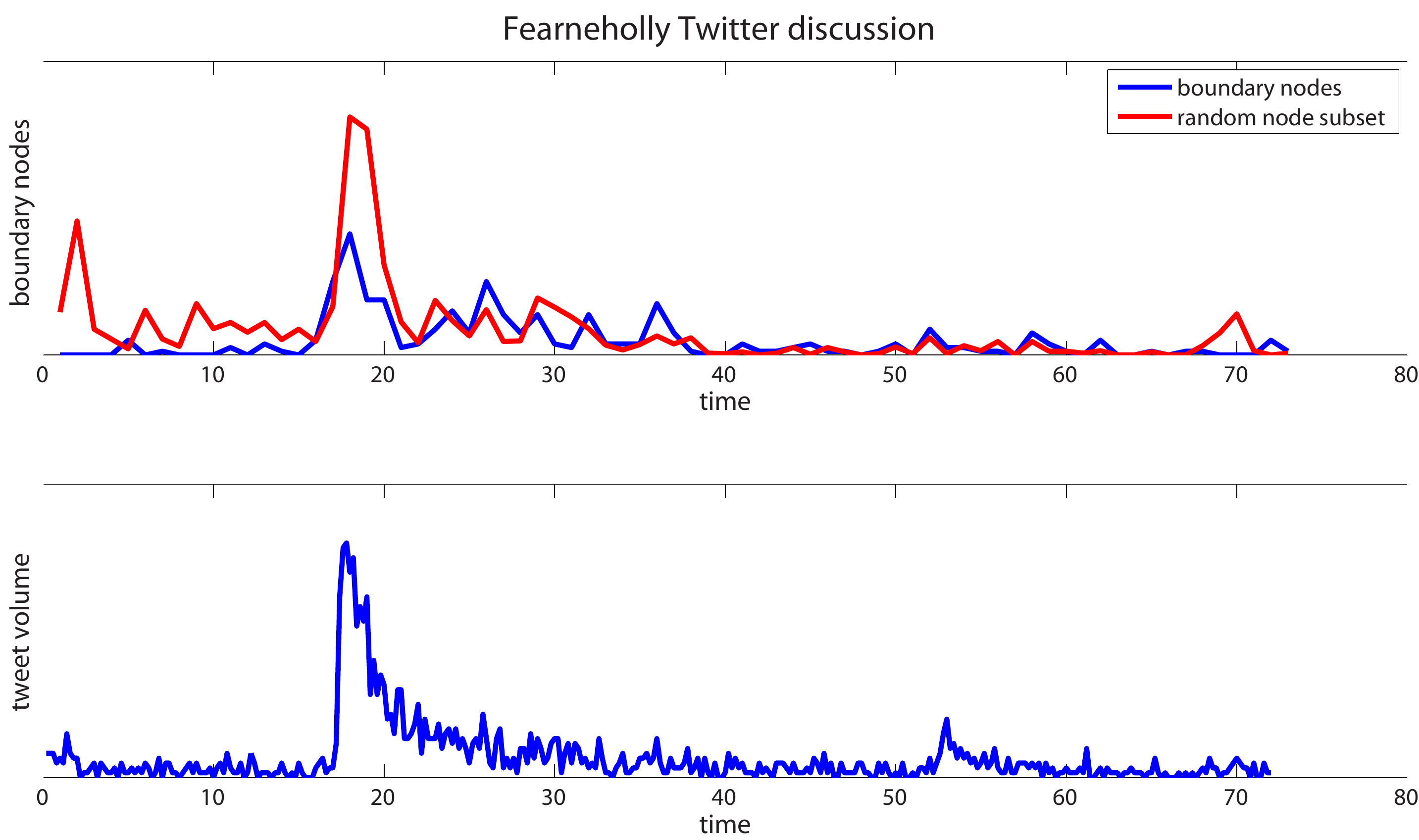}
\caption{\label{fig:temporal1}\csentence{Boundary node detection over time from real time monitoring of events in Twitter} 
The Twitter activity of a TV show is monitored and the Tweets are gathered.
The top plot shows the number of boundary nodes that produce a Tweet between time points in blue, and
in red is shown the number of Tweets from a random set of nodes of equal size to the boundary node set of the Twitter network.
In the bottom plot is the total volume of Tweets over time and we see a single dominant spike in the conversation activity with a decay trailing afterwards.
We can see that the boundary nodes produce a single dominant spike mirroring that of the total conversation activity.
The random set of nodes is selected to see whether the boundary node activity simply reflects the total number of Tweets, but we can see a spike in the activity occurs at the start of the monitoring which is not present in either of the other trajectories. Since this spike is not present in the total volume trajectory it can be seen as a false positive and the boundary node subset is a better indicator and predictor for conversation intensity.}
\end{center}
\end{figure}

\section*{Tables}

\begin{table}[h!]
\label{alg:outline}
\textbf{Outline of the boundary node vicinity algorithm}
\begin{enumerate}
\item extract the set of connected graphs from the original graph
\item for each connected component obtain the community labels for the graph
\item obtain the set of boundary nodes
\item measure the local vicinity of each boundary node using the fixed length random walk method and aggregate all of the values in the graph into a normalised score
\end{enumerate}
\end{table}

\end{backmatter}
\end{document}